\documentstyle[prb,aps,twocolumn]{revtex}

\begin{document}
\draft \flushbottom

\twocolumn[\hsize\textwidth\columnwidth\hsize\csname
@twocolumnfalse\endcsname

\title{Angle-resolved photoemission study and first principles
calculation of the electronic structure of GaTe}

\author{J.F. S\'{a}nchez-Royo}
\address{ICMUV, Univ. de Valencia, c/Dr. Moliner 50, 46100
Burjassot, Valencia, Spain\\
LURE, Centre Universitaire Paris-Sud,
B\^{a}t. 209 D, B.P. 34, 91898 Orsay Cedex, France}

\author{J. Pellicer-Porres, A. Segura, and V. Mu\~{n}oz-Sanjos\'{e}}
\address{ICMUV, Univ. de Valencia, c/Dr. Moliner 50, 46100
Burjassot, Valencia, Spain}

\author{G. Tob\'{\i}as, P. Ordej\'{o}n, and E. Canadell}
\address{ICMAB, CSIC, Campus de la Univ. Aut\`{o}noma de
Barcelona, 08193 Bellaterra, Barcelona, Spain}

\author{Y. Huttel}
\address{LURE, Centre Universitaire Paris-Sud, B\^{a}t. 209 D,
B.P. 34, 91898 Orsay Cedex, France}

\date{\today}
\maketitle
\begin{abstract}

The electronic band structure of GaTe has been calculated by
numerical atomic orbitals density-functional theory, in the local
density approximation. In addition, the valence-band dispersion
along various directions of the GaTe Brillouin zone has been
determined experimentally by angle-resolved photoelectron
spectroscopy. Along these directions, the calculated valence-band
structure is in good concordance with the valence-band dispersion
obtained by these measurements. It has been established that GaTe
is a direct-gap semiconductor with the band gap located at the Z
point, that is, at Brillouin zone border in the direction
perpendicular to the layers. The valence-band maximum shows a
marked \textit{p}-like behavior, with a pronounced anion
contribution. The conduction band minimum arises from states with
a comparable \textit{s}- \textit{p}-cation and \textit{p}-anion
orbital contribution. Spin-orbit interaction appears to specially
alter dispersion and binding energy of states of the topmost
valence bands lying at $\Gamma$. By spin-orbit, it is favored
hybridization of the topmost \textit{p}$_z$-valence band with
deeper and flatter \textit{p$_x$}-\textit{p$_y$} bands and the
valence-band minimum at $\Gamma$ is raised towards the Fermi level
since it appears to be determined by the shifted up
\textit{p$_x$}-\textit{p$_y$} bands.

\end{abstract}

\pacs{PACS numbers: 71.20.Nr, 79.60.Bm} ]


\section{Introduction}

Layered materials of the III-VI family like GaSe, InSe, and GaTe
are of special interest for their potential applications in
nonlinear and optical bistable devices,\cite{Eckhoff,Mangeles} as
well as in the development of solar cells and solid-state
batteries.\cite{juanpastor,juan,bateria1,bateria2} Among these
compounds, GaTe is one of the less studied because it presents a
higher anisotropic and more complex crystal structure than the
rest of the III-VI materials. GaTe belongs to the B2/m space group
and has a base centered monoclinic unit cell.\cite{Julien} The
layer structure [see Fig.~\ref{fig1}(a)], as that of GaSe or InSe,
is composed of a four-sheet anion-cation-cation-anion intralayer
stacking pattern in which the bonds are mainly covalent with some
ionic contribution. Nevertheless, in contrast to GaSe and InSe,
one-third of the cation-cation bonds in GaTe lies almost in the
layer plane in a direction perpendicular to the \textit{c} axis.
Differences also exist when considering layer stacking sequences.
Similarly to the rest of III-VI compounds, interlayer anion-anion
bonds in GaTe are of weak van der Waals type. Generally, this fact
promotes the existence of polytypes with different stacking
sequences, as is observed in InSe and
GaSe.\cite{politipos1,politipos2} Nevertheless, no polytypism
seems to occur in GaTe, as is suggested by the absence of
conjugate modes in Raman-scattering and infrared-absorption
measurements.\cite{Irwin} The whole structure gives rise to
fourfold coordination for the Ga atoms, with three Te and one Ga
atoms, and to threefold coordination for the Te with Ga atoms. As
a consequence of these differences in the crystal structure of
GaTe with respect to other layered compounds, GaTe has only one
twofold symmetry axis, lying along the \textit{c} axis, whereas
other III-VI compounds have a threefold symmetry axis
perpendicular to the layer plane.

Besides the common in-plane-out-of-plane anisotropy observed in
the optical and electrical properties of the layered compounds,
the special in-plane anisotropy in bonding in GaTe makes the
optical properties also anisotropic in the layer plane. GaTe is a
direct-gap semiconductor with strong excitonic absorption even at
room temperature.\cite{opt1,opt2,opt3,opt4} The optical properties
near the band-gap energy region appear to be slightly
anisotropic.\cite{opt2} Anisotropy of the absorption coefficient
has been observed to increase at higher energies, which was
attributed to transitions from deep states related to the in-plane
Ga-Ga bonds.\cite{opt4} Moreover, low-temperature
photoluminescence and transmission measurements revealed the
presence of different exciton states depending on the in-plane
polarization vector.\cite{wan} Recently, these results were
explained by the existence of three types of exciton states due to
\textit{j-j} coupling, whose optical transitions are allowed under
the group symmetry of this material.\cite{Yamamoto}

There is still a lack of both theoretical and experimental studies
on the electronic band structure of GaTe. One of the reasons for
this lacuna lies on the complexity of the primitive unit cell,
that contains 6 GaTe units (54 valence electrons), which has made
band structure calculations hardly available up to now. Recently,
the electronic band structure of GaTe has been calculated along
some directions by the \textit{ab initio} tight-binding linear
muffin-tin orbitals method.\cite{Yamamoto} These results are in
agreement with the direct-gap behavior of this semiconductor and
locate the band edge at the border of the Brillouin zone (BZ).
Nevertheless, there are still some open questions regarding the
electronic band structure along the high-symmetry directions of
the GaTe BZ, as well as the experimental confirmation of the
electronic band structure obtained in the calculations.

In this work we study the electronic properties of the GaTe
compound. The GaTe band structure has been calculated by numerical
atomic orbitals density-functional theory (NAO-DFT). In addition,
angle-resolved photoemission (ARPES) measurements have been
carried out in GaTe(001) single crystals. From these measurements,
the valence-band dispersion of GaTe was extracted and interpreted
on the basis of the band structure calculations. The experimental
setup is described in Sec. II. Section III is devoted to the band
structure of GaTe calculated along different directions of the
GaTe BZ, including the technical aspects of the calculation. In
Sec. IV, the experimental results obtained by ARPES are presented
and discussed in the frame of the calculated GaTe band structure.
Finally, in Sec. V we present the main conclusions of this work.

\section{Experimental}

The experiments were performed at LURE (Orsay, France) using the
Spanish-French (PES2) experimental station of the Super-Aco
storage ring, described elsewhere.\cite{campana} The measurements
were carried out in a purpose-built ultra-high vacuum system, with
a base pressure of 5x10$^{-11}$ mbar, equipped with an angle
resolving 50 mm hemispherical VSW analyzer coupled on a goniometer
inside the chamber. The manipulator was mounted in a two-axes
goniometer which allows rotation of the sample in the whole
360$^o$ azimuthal angle and in the 180$^o$ polar emission angle
relative to surface normal ($\Theta _{off}$), with an overall
angular resolution of 0.5$^o$. Photoelectrons were excited with
p-polarized synchrotron radiation in the 20-60 eV energy range.
For this selected photon energy (h$\nu$) range, the energy
resolution was of 60 meV. With this setup, different
energy-distribution-curve (EDC) series were recorded scanning the
h$\nu$ as well as $\Theta _{off}$, at constant incident angle of
the light ($\Theta _{i}$). In these conditions, changes of
polarization effects on initial states along one series are
neglected. All the EDC spectra shown in this work are referred to
the Fermi level (E$_F$).

The GaTe single crystals used in this work were cut from an ingot
grown by the Bridgman-Stockbarger method. The samples were cleaved
\textit{in situ} after introduction in the ultra-high vacuum
chamber. The samples were easily cleaved in the layer plane due to
the existence of weak interlayer van der Waals bonds. LEED spectra
showed a sharp spot pattern corresponding to the bulk monoclinic
material. No surface impurities were detected by synchrotron
radiation photoemission measurements. The samples were oriented by
azimuthal and polar photoelectron diffraction scans recording the
Ga 3\textit{d} peak intensity. Unambiguous azimuthal orientation
of the samples was possible due to the B2/m symmetry of the
crystal.

\section{Band structure of GaTe}

\subsection{Calculation details}

In this work, we present fully self-consistent density-functional
theory\cite{DFT} (DFT) calculations of the electronic structure of
GaTe. The calculations are performed in the local-density
approximation.\cite{LDA} The exchange-correlation potential is
that of Ceperley and Adler\cite{Adler} as parameterized by Perdew
and Zunger.\cite{Zunger} Only the valence electrons are considered
in the calculation, with the core being replaced by
norm-conserving scalar relativistic pseudopotentials\cite{pseudo}
factorized in the Kleinman-Bylander form.\cite{Kleinman} The
pseudopotentials were generated using the atomic valence
configurations 5\textit{s}$^2$5\textit{p}$^4$ for Te and
3\textit{d}$^{10}$4\textit{s}$^2$4\textit{p}$^1$ for Ga. The
cutoff radii were 2.0, 2.0, 3.0, and 3.0 a.u. for the \textit{s},
\textit{p}, \textit{d}, and \textit{f} components in Te,
respectively, and 2.1, 2.3, 1.1 and 2.0 a.u. for the \textit{s},
\textit{p}, \textit{d}, and \textit{f} components of Ga,
respectively. Since the core and valence charges overlap
significantly for both atomic species, we include nonlinear
partial-core corrections\cite{core} with matching radii of 1.0
a.u. for Te and 0.5 a.u. for Ga to describe the exchange and
correlation in the core region.

The valence one-particle problem was solved using a linear
combination of numerical (pseudo-) atomic orbitals with finite
range.\cite{orbitals} The details of the basis generation
(including multiple-$\zeta$ and polarization functions) can be
found elsewhere.\cite{Artacho} In this work we have used a
split-valence double-$\zeta$ basis set with a single shell of
polarization orbitals (that is, containing two \textit{s} shells,
two \textit{p} shells, and one \textit{d} shell both for Te and
for Ga), as obtained with an energy shift of 250 meV and a split
norm of 15\%.\cite{Artacho} The integrals of the self-consistent
terms of the Kohn-Sham Hamiltonian are obtained with the help of a
regular real space grid on which the electron density is
projected. The Hartree potential is calculated by means of fast
Fourier transforms in that grid. The grid spacing is determined by
the maximum kinetic energy of the plane waves that can be
represented in that grid. In the present work, we used a cutoff of
around 300 Ry (which changes slightly with the volume of the unit
cell). A regular grid of 5x5x5 \textit{k} points\cite{matrix} was
used to sample the BZ. We have checked that the results are well
converged with respect to the real space grid, the BZ sampling and
the range of the numerical atomic orbitals. The calculations were
performed using the {\sc Siesta} code.\cite{siesta}

In our band structure calculations, we used a centered monoclinic
cell with primitive vectors
$\textbf{a$'$}=-\textbf{b}+(\textbf{c}-\textbf{a})/2$,
$\textbf{b$'$}=-\textbf{b}-(\textbf{a}+\textbf{c})/2$, and
$\textbf{c$'$}=-\textbf{b}$, where \textbf{a}, \textbf{b}, and
\textbf{c} are the repeat vectors of the monoclinic cell [see
Figure~\ref{fig1}(a)]. Note that, with this election,
\textbf{a$'$} and \textbf{b$'$} vectors are in the layer plane.

\subsection{Brillouin zone and electronic band structure}

Figure~\ref{fig1}(a) shows the crystal structure of GaTe,
described previously. In this figure, we have indicated the
crystallographic axes of the GaTe monoclinic unit cell. The GaTe
crystal structure is quite similar to those of GaSe and InSe,
except for the existence of cation-cation bonds lying almost in
the layer plane, with the same coordination and bonding behavior.
The comparison can be also extended to the reciprocal space.
Figure~\ref{fig1}(b) shows the calculated GaTe BZ, in which the
main points are indicated. The relation between the different
symmetry directions of the GaTe BZ and the reciprocal lattice
vectors of the centered monoclinic unit cell is shown in
Table~\ref{tab:table1}.

The GaTe BZ can be easily described in terms of the crystal
structure. The binary axis corresponds to the $\Gamma$Y
high-symmetry direction and a crystal vector perpendicular to the
layer plane defines a direction parallel to the $\Gamma$Z
high-symmetry direction, in the reciprocal space. The GaTe BZ
resembles that of InSe, but contracted along the $\Gamma$X
direction. The contraction is due to the presence of the
\textit{quasi}-in-plane Ga-Ga bonds in GaTe, which produces that
the fundamental translation in the layer plane consists of three
Ga$_2$Te$_2$ unities, in contrast to one unity in the other III-VI
compounds. On the other side, as both GaTe and InSe have only one
layer per unit cell, one should expect the electronic band
structure of GaTe to be related to that of InSe, after a
three-folding in the layer plane.

Figure~\ref{fig2} shows the electronic band structure of GaTe
along different directions of the GaTe BZ [Fig.~\ref{fig1}(b)]
calculated by the NAO-DFT method described previously. The origin
of binding energy has been taken at the E$_F$. In spite of the
large number of bands obtained, a detailed description of the
valence band can be done. The GaTe unit cell consists of three
Ga$_2$Te$_2$ unities. Therefore, there are per unit cell: three
Ga-Ga bonds, 18 Ga-Te bonds, and one electronic pair associated to
each one of the six Te. This gives rise to 27 doubly occupied
bands, which are precisely those appearing with binding energies
between 14 eV and E$_F$ and composing the GaTe valence band. Below
this energy (at 16.5 eV), the 3\textit{d}-core levels of Ga are
present (not shown). The unoccupied side of the band structure is
rather more complicated to be described, since, besides the 18
Ga-Te and 3 Ga-Ga antibonding levels, all contributions from the
polarization \textit{d}-orbitals overlap.

A detailed band to band description of the band structure of GaTe
would not be very useful due to the fact that the low symmetry of
the crystal makes the orbital mixture of each band to be very
high. Nevertheless, a general description of the orbital character
of the GaTe band structure can be performed by means of the
density of states (DOS). Figure~\ref{fig3}(a) shows the total DOS
calculated by the NAO-DFT method. Figure~\ref{fig3}(b)
and~\ref{fig3}(c) show the different projections of the
\textit{s}, \textit{p}, and \textit{d} orbitals onto the DOS for
Ga and Te, respectively. These results indicate that the valence
band of GaTe can be divided into three different groups: (i) One
group of six bands appearing between 14 and 11 eV, which show a
pronounced Te 5\textit{s} relative behavior with a small Ga
4\textit{s} and 4\textit{p} contribution, (ii) a second group of
six bands appearing between 8 and 5 eV, which mostly show a Ga
4\textit{s} character with a pronounced Te 5\textit{p}
contribution in the low binding energy side, and (iii) a third
group of 15 bands comprising the range of energy between 5 eV and
the E$_F$, which are a mixture of Te 5\textit{p} and Ga
4\textit{p} orbitals with a small Ga 4\textit{d} and 4\textit{s}
contribution. The origin of these groups differs from one to the
other. On one side, the first and second group of bands appear to
be mostly originated by the Ga-Te bonds, showing those with a
marked \textit{p}-like behavior a more pronounced dispersion along
the $\Gamma$Z high-symmetry direction (as well as along directions
perpendicular to it) than those with a predominant \textit{s}-like
behavior (Fig.~\ref{fig2}). On the other side, in the third group
of bands, they coexist six bands originated also from the Ga-Te
bonds, three bands from the Ga-Ga bonds, and the six electronic
pairs associated to the Te (mostly Te 5\textit{p}). Focusing our
attention on the topmost valence bands, the fact that their
orbital character has an important \textit{p}-like component
perpendicular to the layers, makes that most of the bands display
a noticeable dispersion along $\Gamma$Z high-symmetry direction
(Fig.~\ref{fig2}).

With regard to the DOS corresponding to the unoccupied side of the
band structure, the huge overlap of Ga-Te and Ga-Ga antibonding
levels with the \textit{d}-orbitals of polarization makes an
orbital description quite more complicated than that carried out
for the occupied side of the DOS. Nevertheless, it should be
mentioned that states at the conduction band edge are mostly
originated from Te 5\textit{p} and Ga 4\textit{p} orbitals, with
an important contribution from Ga 4\textit{s} and a smaller
contribution Te 5\textit{d} and Ga 4\textit{d}, which suggests a
pronounced \textit{sp$_z$}-like behavior to the conduction band
edge. Only for higher states in the conduction band, contributions
from Ga and Te \textit{d}-orbitals of polarization appear to be of
the same magnitude than those of the \textit{p}-orbitals.

The general shape and the different orbital contributions to the
calculated DOS by NAO-DFT are quite similar to those obtained
previously by the \textit{ab initio} tight binding
method.\cite{Yamamoto} Some differences can be appreciated in the
unoccupied side of the DOS. By both methods, the behavior of the
conduction band edge seems to be established as a mixture of
\textit{s} and \textit{p} orbitals. Nevertheless, in contrast to
previous calculations,\cite{Yamamoto} the Ga 4\textit{p}
contribution to the conduction band edge appears to be of the same
order as that of the Ga 4\textit{s}, which gives a more pronounced
\textit{p}-like behavior to the conduction band edge in the
present calculations. Also the contribution of Ga and Te
\textit{d}-orbitals of polarization to the conduction band edge,
as well as to deeper states of the conduction band, seem to be
underestimated in those calculations (mostly the Ga 4\textit{d}).
As regards the band dispersion, authors seem to have used a
standard six-face monoclinic BZ,\cite{Yamamoto} which does not
correspond to the GaTe BZ. The band dispersion calculated along
some directions in Ref.~\onlinecite{Yamamoto} (see
Table~\ref{tab:table1}) is very similar to the present results.
However, along other directions, as the $\Gamma$Z and AM
directions in Fig. 8 of Ref.~\onlinecite{Yamamoto}, the path
crosses out of the first BZ to neighboring ones, giving rise to
the \textit{quasi}-symmetrical band dispersion observed along
these directions.

From the above results, it can be concluded that GaTe is a
direct-gap semiconductor with the gap located at zone border at Z
point of the GaTe BZ, which has the same symmetry as $\Gamma$
(B2/m). The band gap appears to be underestimated (1.098 eV),
which is a well-known tendency of the LDA approximation in
semiconductors. As in InSe,\cite{manjon} the valence-band maximum
is mostly formed by antibonding \textit{p$_z$}-orbitals of the
anion. Nevertheless, in contrast to InSe,\cite{manjon} the
conduction-band minimum is not only formed by the
\textit{s}-orbital contribution of the cation. In GaTe a more
intense contribution of cation and anion \textit{p}-orbitals can
be observed. This can be attributed to the fact that the presence
of the \textit{quasi}-in-plane cation-cation bonds distorts the
anion disposition and, therefore, the interaction of Te atoms with
those of contiguous layers and with Ga atoms increases.

\section{Experimental Results and discussion}

In this section, we compare the calculated electronic band
structure of GaTe with the band dispersion determined
experimentally by ARPES in GaTe(001) single crystals. First, we
will show and discuss the electronic band dispersion along the
$\Gamma$Z high-symmetry direction, proceeding later with those
determined along different directions perpendicular to it.

Figure~\ref{fig4}(a) shows an EDC series recorded at normal
emission with constant $\Theta _{i}$ in a GaTe(001) sample, as a
function of the h$\nu$. In these spectra, several peaks have been
identified and labelled by small solid bars. In some of them, two
sharp dispersing peaks can be also observed, which can be
attributed to second-order transitions from deeper occupied states
induced by the experimental system. As the h$\nu$ increases, these
sharp peaks move to lower binding energies. Therefore,
photoemission signal from deeper valence-band transitions than
those of Fig.~\ref{fig4}(a) can be studied in EDCs recorded by
extending up the h$\nu$-range of Fig.~\ref{fig4}(a).
Figure.~\ref{fig4}(b) shows the results of these measurements. In
these spectra, only broad features can be identified, at binding
energies of 12.6 eV, which have been labelled as \textit{s-band}.

By these results, the dispersion of states along the $\Gamma$Z
high-symmetry direction can be determined. Nevertheless, to
compare these results with the calculated band structure, the
momentum of photoelectrons should be estimated. Photoemission
process involves energy and crystal momentum conservation. In the
frame of the three-step model of the photoexcitation mechanism in
ARPES, both momentum components, \textit{k$_\parallel$} and
\textit{k$_\perp$}, of the photoelectrons inside a bulk material
can be expressed as\cite{Hufner}

\begin{mathletters}
\label{ec1}
\begin{equation}
k_{\Vert }=\sqrt{\frac{2m}{\hbar ^2}}\sqrt{h\nu-BE-\Phi }\sin
\left(\Theta _{off}\right)\label{ec1:1}
\end{equation}
and
\begin{equation}
k_{\bot }=\sqrt{\frac{2m}{\hbar ^2}}\sqrt{\left( h\nu-BE-\Phi
\right) \cos ^2\left( \Theta _{off}\right) -V_o}\;,\label{ec1:2}
\end{equation}
\end{mathletters}
where \textit{BE} is the binding energy, $\Phi$ is the work
function, \textit{V$_o$} is the inner potential, \textit{m} is the
free-electron mass, and \textit{$\hbar $} is the reduced Planck
constant. A complete determination of the photoelectron momentum
requires an accurate estimate of the \textit{V$_o$}
[Eq.(\ref{ec1:2})]. We have assumed it as a constant parameter. An
initial testing value of \textit{V$_o$}$\sim$-19 eV ($\Phi$$\sim$6
eV) can be obtained from the bottom of the valence band
(Fig.~\ref{fig2}). After testing several values of \textit{V$_o$}
and taking into account that k$_\Gamma$$_Z$=0.421 \AA$^{-1}$, we
have adopted a value of \textit{V$_o$}=-25.7 eV, which is two
times higher than that found for InSe.\cite{InSeband} Such a high
value of the \textit{V$_o$} for GaTe is not surprising, since the
orbital hybridization giving rise to the GaTe valence-band
structure is higher than in InSe. In any case, it is reasonably
close to the value expected from the valence-band calculations.

Figure~\ref{fig4}(c) shows the valence-band dispersion extracted
from the peaks identified in Fig.~\ref{fig4}(a), by means of
Eqs.~(\ref{ec1}). The calculated valence-band dispersion along the
$\Gamma$Z high-symmetry direction has been also included. In spite
of the large number of different bands present, it can be observed
that the calculated band structure reproduces most of the traces
observed in the experimental band dispersion. The large
concentration of quite flat bands appearing at lower binding
energies is well reproduced by the calculated valence-band
structure, with states lying at similar binding energies. These
facts also put forward the highly hybridized behavior of top
states of the valence-band structure predicted by calculations
(Fig.~\ref{fig2}). Besides this, dispersion of states with a
marked Te 5\textit{p}-Ga 4\textit{s} orbital character (that is,
states between 8 and 5 eV in binding energy) is quite well
reproduced. Moreover, the presence of the broad \textit{s-band}
peaks in Fig.~\ref{fig4}(b) can be attributed to transitions from
the Te 5\textit{s}-like states, located at binding energies of
$\sim$12 eV (Fig.~\ref{fig2}). Nevertheless, some discrepancies
appear for states lying at the top of the valence band.
Experimentally, the valence-band maximum along the $\Gamma$Z
high-symmetry direction appears to be at Z point and the minimum
of the valence-band lies at $\Gamma$. These facts are in
concordance with the calculated valence-band structure. Moreover,
dispersion of the topmost band around Z is well reproduced by the
calculations. However, the valence-band minimum at $\Gamma$
obtained by calculations appears shifted down by 0.8 eV with
respect to the experimental one. A similar situation occurs with
the second topmost valence band. For this band, the calculated
band minimum, also at $\Gamma$, appears to be shifted down by 0.6
eV with respect to the experimental one. This disagreement will be
discussed later on.

Figure~\ref{fig5}(a) shows an EDC series measured by ARPES with
h$\nu$=33 eV in a GaTe(001) sample. The sample orientation has
been chosen to probe, by these ARPES measurements, points of the
reciprocal space contained in the $\Gamma$ZH plane. In fact, for
this selected h$\nu$, these points are expected to have a
k$_{\perp}$ lying quite close to the Z point [Ecs.(\ref{ec1})]. In
the spectra shown in Fig.~\ref{fig5}(a), several peaks have been
identified and labelled by small solid bars. Figure~\ref{fig5}(b)
shows the band dispersion extracted from the peaks identified in
Fig.~\ref{fig5}(a).

In order to compare the experimental band dispersion shown in
Fig.~\ref{fig5}(b) with the calculated band structure, it should
be taken into account that, by the ARPES measurements summarized
in Fig.~\ref{fig5}(a), particular points of the reciprocal space
along a parabolic curve are probed for each constant initial state
[Eqs.~(\ref{ec1})]. For initial states with low binding energy,
the minimum of the parabola will be close to the Z point, but when
running from Z to H (k$_{ZH}$=0.744 \AA$^{-1}$) the parabolic
curve perpendicularly deviates from horizontality by 0.07
\AA$^{-1}$. For deeper initial states, the parabolas probe points
somewhere along the $\Gamma$Z high-symmetry direction. These facts
produce that, by these measurements, states along the ZH direction
are not strictly probed. Nevertheless, it has been found that
dispersion of valence-band states along the $\Gamma$Z
high-symmetry direction can be considered negligible in a large
k$_{\perp}$-range [Fig.~\ref{fig4}(c)]. These facts encourages us
to assume that the calculated band structure along the ZH
direction is a good approximation for the experimental band
dispersion, as shown in Fig.~\ref{fig5}(b). In this figure, it has
been also introduced the calculated band dispersion of states with
binding energies higher than 3 eV along the $\Gamma$Y
high-symmetry direction, which is not far from that calculated
along the ZH direction and gives a better agreement with the
experimental band dispersion of deeper bands.

Figure~\ref{fig6}(a) and~\ref{fig6}(b) show two different EDC
series measured by ARPES with h$\nu$=33 eV in a GaTe(001) sample.
In these ARPES measurements, a sample orientation has been chosen
to probe points of the reciprocal space contained in the
$\Gamma$ZM plane. In these spectra, several peaks have been
identified and labelled by small solid bars. Figure~\ref{fig6}(c)
shows the $\Gamma$ZM transversal cut of the GaTe BZ, in the
extended zone scheme. The parabolic curve defined by ARPES
measurements with h$\nu$=33 eV for constant initial states at the
E$_F$ is also included in this figure. This scheme illustrates the
fact that, by both sets of measurements shown in
Figs.~\ref{fig6}(a) and~\ref{fig6}(b), points of the reciprocal
space are probed along the MZM' and M$\Gamma$M' directions,
respectively. Figure~\ref{fig6}(d) shows the band dispersion
extracted from the peaks identified in Figs.~\ref{fig6}(a)
and~\ref{fig6}(b). Assuming the same hypotheses as those discussed
previously, we have also included the calculated band dispersion
along the MZM' and M$\Gamma$M' directions, which reproduce most of
the traces of the band dispersion obtained experimentally.
Focusing again attention on the two topmost valence-bands, their
dispersion around Z along the MZM' direction appears to be rather
flat, coherently with the behavior expected by NAO-DFT
calculations. Nevertheless, dispersion of these bands along the
M$\Gamma$M' direction shows the same behavior as that observed
along the $\Gamma$Z high-symmetry direction [Figs.~\ref{fig4}(c)],
with the same separation at $\Gamma$ between the calculated and
experimental minimums of these bands.

The above results obtained by ARPES measurements along different
directions of the GaTe BZ support the results obtained by NAO-DFT
band structure calculations (Fig.~\ref{fig2}). Among these
results, it should be emphasized the fact that experimental and
calculated band structure place the valence-band maximum in GaTe
at zone border at Z, similarly as occurs in the InSe
compound.\cite{manjon} Nevertheless, despite the good agreement
obtained through the whole reciprocal space studied, there still
remains one point of discrepancy related to the two topmost
valence-bands at $\Gamma$ [Figs.~\ref{fig4}(c) and~\ref{fig6}(d)],
since the binding energy of the minimum of these bands appear to
be overestimated by calculations. These facts can be explained
since spin-orbit interaction was not considered in the present
calculations. In the related compound InSe, the band giving rise
to the valence band maximum at Z mostly has a marked
\textit{p$_z$}-orbital character. States of this band have a total
orbital moment \textit{j}=1/2. Without considering spin-orbit
interaction, this band disperses down along the $\Gamma$Z
direction, showing the minimum at $\Gamma$. Below this band, the
next ones are quite flat bands with a
\textit{p$_x$}-\textit{p$_y$} character. Spin-orbit interaction
splits off \textit{p$_x$}-\textit{p$_y$} states with different
\textit{j}, rising those with \textit{j}=3/2 by 0.3-0.4
eV,\cite{bandstructure} with respect to the valence band maximum.
This produces that the minimum of the topmost valence band at
$\Gamma$ is determined by the shifted up
\textit{p$_x$}-\textit{p$_y$} bands. A more intense spin-orbit
coupling can be expected in GaTe, since the nature of its topmost
valence bands is the same as those of InSe and its anion has a
higher mass. In fact, the spin-orbit induced shift on the
valence-band minimum at $\Gamma$ in GaTe may be evaluated. Taking
into account that the atomic spin-orbit
\textit{p}$_{3/2}$-\textit{p}$_{1/2}$ splitting is of 589 and 247
meV for Te and Se, respectively,\cite{spinorbit} in GaTe one
should expect a shift up of the \textit{j}=3/2
\textit{p$_x$}-\textit{p$_y$} bands at $\Gamma$ of the order of
0.7-0.9 eV, which is close to that observed in Figs.~\ref{fig4}(c)
and~\ref{fig6}(d).

\section{Conclusions}

The electronic band structure of GaTe single crystal has been
calculated by a NAO-DFT method. These results show that, on one
side, the deepest valence bands are mostly formed by Ga-Te
orbitals with a marked \textit{s}-like behavior and, on the other
side, bands with a low binding energy have a pronounced
\textit{p}-orbital character originated by cation-anion bonds, as
well as by anion electronic pairs and cation-cation bonds. Also,
the conduction-band structure was determined. It appears to be
composed by bands with mainly a \textit{p}-orbital character,
originated from the antibonding Ga-Te and Ga-Ga bonds, intermixed
with the d-orbitals of polarization.

The electronic properties of GaTe single crystal have been studied
also by ARPES measurements. By these measurements, dispersion of
the valence bands along various directions of the GaTe BZ has been
determined. The experimental band dispersion obtained appears to
be quite well reproduced by the calculated band dispersion along
these directions.

These results establish that GaTe is a direct-gap semiconductor
with a gap located at zone border, at Z point. The valence-band
maximum shows a marked \textit{p}-like behavior, with a pronounced
anion contribution. The conduction band minimum arises from states
with a comparable \textit{s}- \textit{p}-cation and
\textit{p}-anion orbital contribution. Spin-orbit interaction
appears to specially alter the dispersion and binding energy of
states of the topmost valence bands lying at $\Gamma$. Spin-orbit
favors the hybridization of the topmost \textit{p}$_z$-valence
band with deeper and flatter \textit{p$_x$}-\textit{p$_y$} bands.
Also, it rises the valence-band minimum at $\Gamma$ towards the
E$_F$, which appears to be determined by the shifted up
\textit{j}=3/2 \textit{p$_x$}-\textit{p$_y$} bands.

\acknowledgments

This work was financed by the Large Scale Facilities program of
the EU to LURE. The authors gratefully acknowledge Dr. J. Avila
and Dr. M.C. Asensio for technical and scientific support at LURE.
J.F.S.-R. and G.T. acknowledge financial support from the
Ministerio de Educaci\'{o}n y Cultura and Ministerio de Ciencia y
Tecnolog\'{\i}a of Spain, respectively. Part of this work was supported
by DGI-Spain (Project BFM2000-1312-C02-01), Generalitat de
Catalunya (Project 1999SGR207) and Fundaci\'{o}n Ram\'{o}n Areces. The
computations were carried out using the resources of CESCA and
CEPBA, coordinated by C$^4$.


\begin{table}
\caption{Main directions of the GaTe Brillouin zone as generated
by the reciprocal lattice vectors. These directions have been
identified with those defined in Ref.~\ref{Yamamoto} and expressed
as a function of the cell vectors (\textbf{G$_\textit{i}$}) there
defined.\label{tab:table1}}
\begin{tabular}{cccc} \multicolumn{2}{c}{This work}
&\multicolumn{2}{c}{Ref.~\ref{Yamamoto} \tablenotemark[1]}\\
 \hline
$\Gamma$Z&\textbf{c$'^*$}/2&$\Gamma$P(M)&
$(\textbf{G$_1$}+\textbf{G$_2$}-\textbf{G$_3$})/2$\\
$\Gamma$A&$-\textbf{b$'^*$}/2$&$\Gamma$L&$(\textbf{G$_2$}
-\textbf{G$_3$})/2$\\
$\Gamma$P&$-(\textbf{a$'^*$}+\textbf{b$'^*$})/2$&$\Gamma$X&
$(\textbf{G$_1$}+\textbf{G$_2$}-2\textbf{G$_3$})/2$\\
$\Gamma$N&$-(\textbf{a$'^*$}+\textbf{b$'^*$}+\textbf{c$'^*$})/2$
&$\Gamma$A&
$-\textbf{G$_3$}/2$\\
$\Gamma$B&$(\textbf{a$'^*$}+\textbf{c$'^*$})/2$&$\Gamma$V&
$\textbf{G$_2$}/2$\\
$\Gamma$A$'$&$\textbf{a$'^*$}/2$&$\Gamma$L&$(\textbf{G$_3$}-
\textbf{G$_1$})/2$\\
\end{tabular}
\tablenotetext[1]{$\textbf{G$_1$}=
\textbf{b$'^*$}+\textbf{c$'^*$}$,
$\textbf{G$_2$}=\textbf{a$'^*$}+\textbf{c$'^*$}$, and
$\textbf{G$_3$}=
\textbf{a$'^*$}+\textbf{b$'^*$}+\textbf{c$'^*$}$.}
\end{table}


\begin{figure}
\caption{(a) Schematic view of the crystal structure of GaTe and
its monoclinic unit cell. Non-equivalent atoms have different
shadowing. Big and small circles correspond to Te and Ga atoms,
respectively. (b) GaTe centered monoclinic Brillouin zone, in
which the main points are indicated. Equivalent Q-points (Q
generic) have been labelled as Q', Q'', and so forth. The
$\Gamma$XY plane is parallel to the layers and the $\Gamma$Y
direction corresponds to the binary axis. The $\Gamma$ZX plane
defines the mirror plane.}\label{fig1}
\end{figure}

\begin{figure}
\caption{ Band structure of GaTe calculated by the NAO-DFT method
along various directions of the GaTe BZ.}\label{fig2}
\end{figure}

\begin{figure}
\caption{(a) Density of states of GaTe. (b)-(c) Projections of the
\textit{s}, \textit{p}, and \textit{d} orbitals onto the total
density of states, for Ga and Te, respectively.}\label{fig3}
\end{figure}

\begin{figure}
\caption{(a)-(b) Normal-emission valence-band EDCs measured in
GaTe(001) single crystal, as a function of the h$\nu$. The h$\nu$
range scanned has been indicated in both figures. The different
peaks identified in the EDCs in (a) have been marked by small
solid bars. The broad peaks identified in the EDCs in (b) have
been labelled as \textit{s-band}. (c) (circles) Band dispersion
diagram extracted from the peaks identified in (a). The calculated
GaTe band structure along the Z$\Gamma$Z high-symmetry direction
has been also plotted, in solid lines.}\label{fig4}
\end{figure}

\begin{figure}
\caption{(a) Valence-band EDCs measured by ARPES with h$\nu$=33 eV
in GaTe(001) single crystal. For this selected h$\nu$, the ZH
direction is scanned by ARPES. The different peaks identified have
been marked by small solid bars. (b) (circles) Band dispersion
diagram extracted from the peaks identified in (a). The calculated
GaTe band structure along the ZH direction has been also plotted,
in solid lines. The calculated band dispersion of states with
binding energies higher than 3 eV along the $\Gamma$Y
high-symmetry direction has been also plotted, in dashed lines. As
a reference, the \textit{k$_\parallel$}-projection of some points
of the GaTe BZ are indicated at the top.}\label{fig5}
\end{figure}

\begin{figure}
\caption{(a) Valence-band EDCs measured by ARPES with h$\nu$=33 eV
in GaTe(001) single crystal. For this selected h$\nu$, the ZM
direction is scanned by ARPES. (b) The same that (a) extending the
$\Theta _{off}$-range to scan the M$\Gamma$M' direction. In both
EDC series, the different peaks identified have been marked by
small solid bars. (c) Schematic view of the BZ cut with the
$\Gamma$ZM plane. The parabolic curve defined by ARPES
measurements with h$\nu$=33 eV for constant initial states at the
E$_F$ is also included. (d) (circles, squares) Band dispersion
diagram extracted from the peaks identified in (a) and (b),
respectively. The calculated GaTe band structure along the MZM'
and M$\Gamma$M' directions has been also plotted, in solid lines.
As a reference, the \textit{k$_\parallel$}-projection of some
points of the GaTe BZ are indicated at the top.}\label{fig6}
\end{figure}

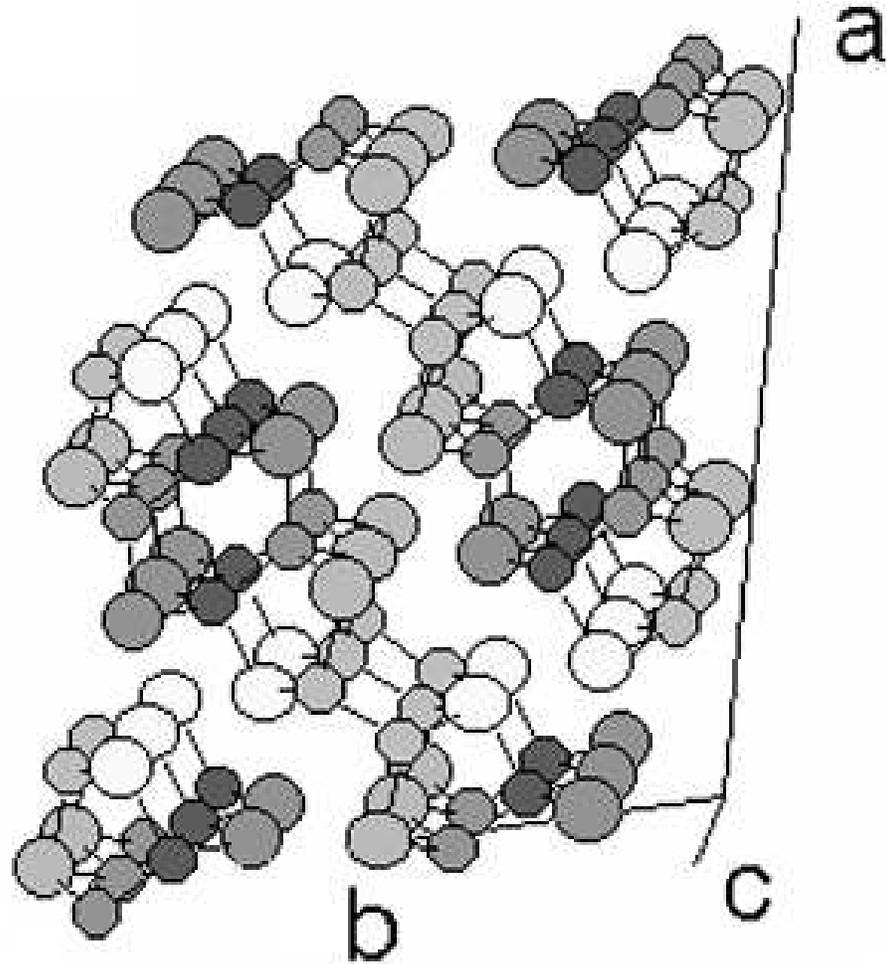
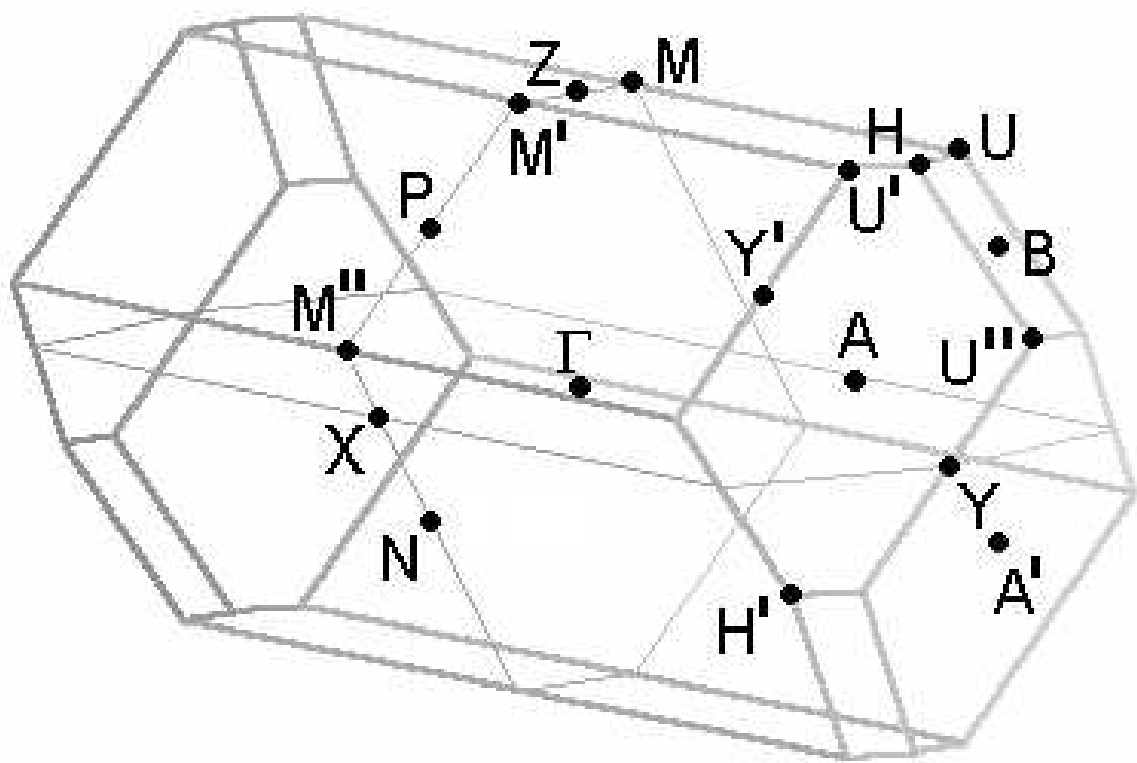
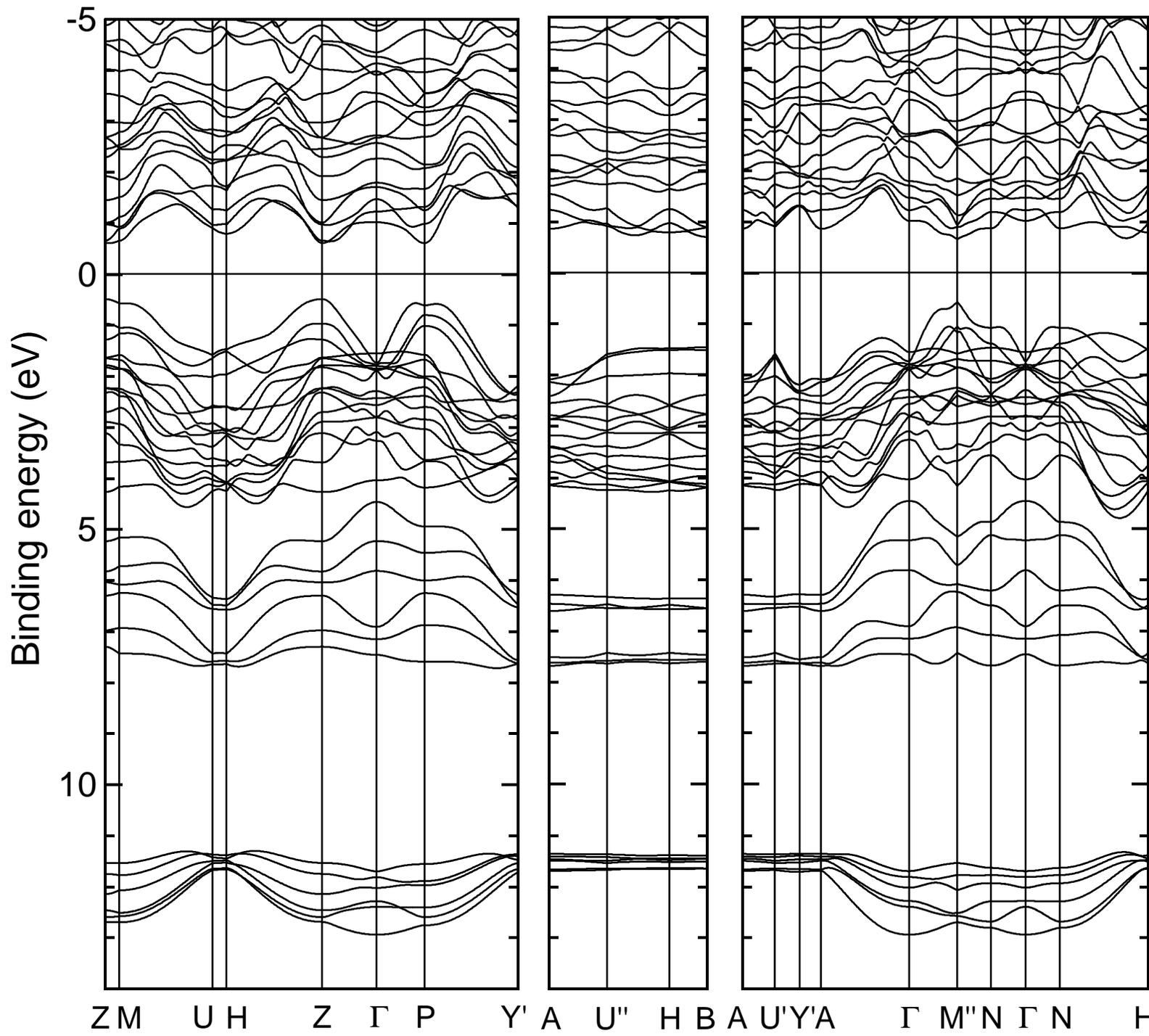
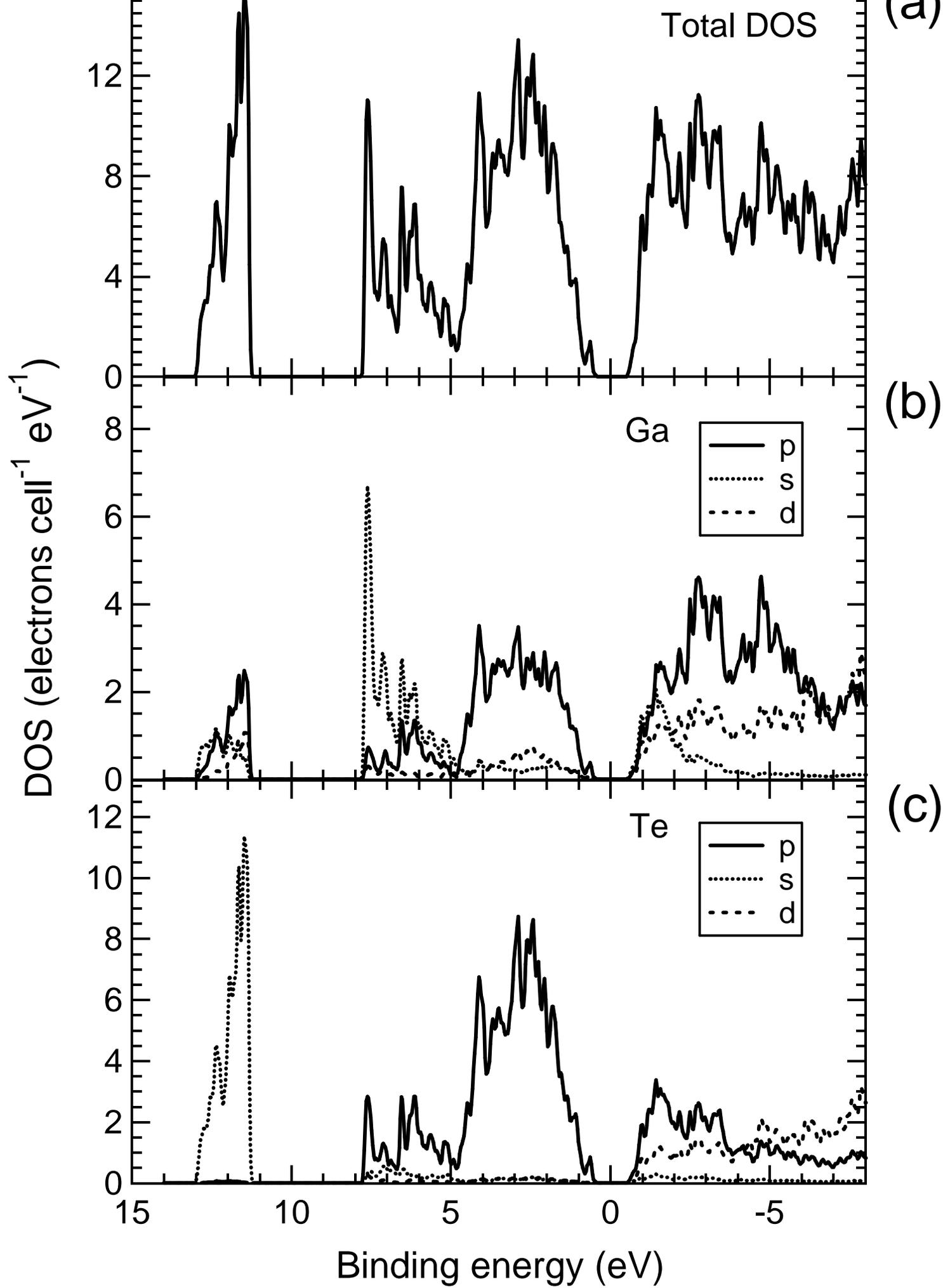
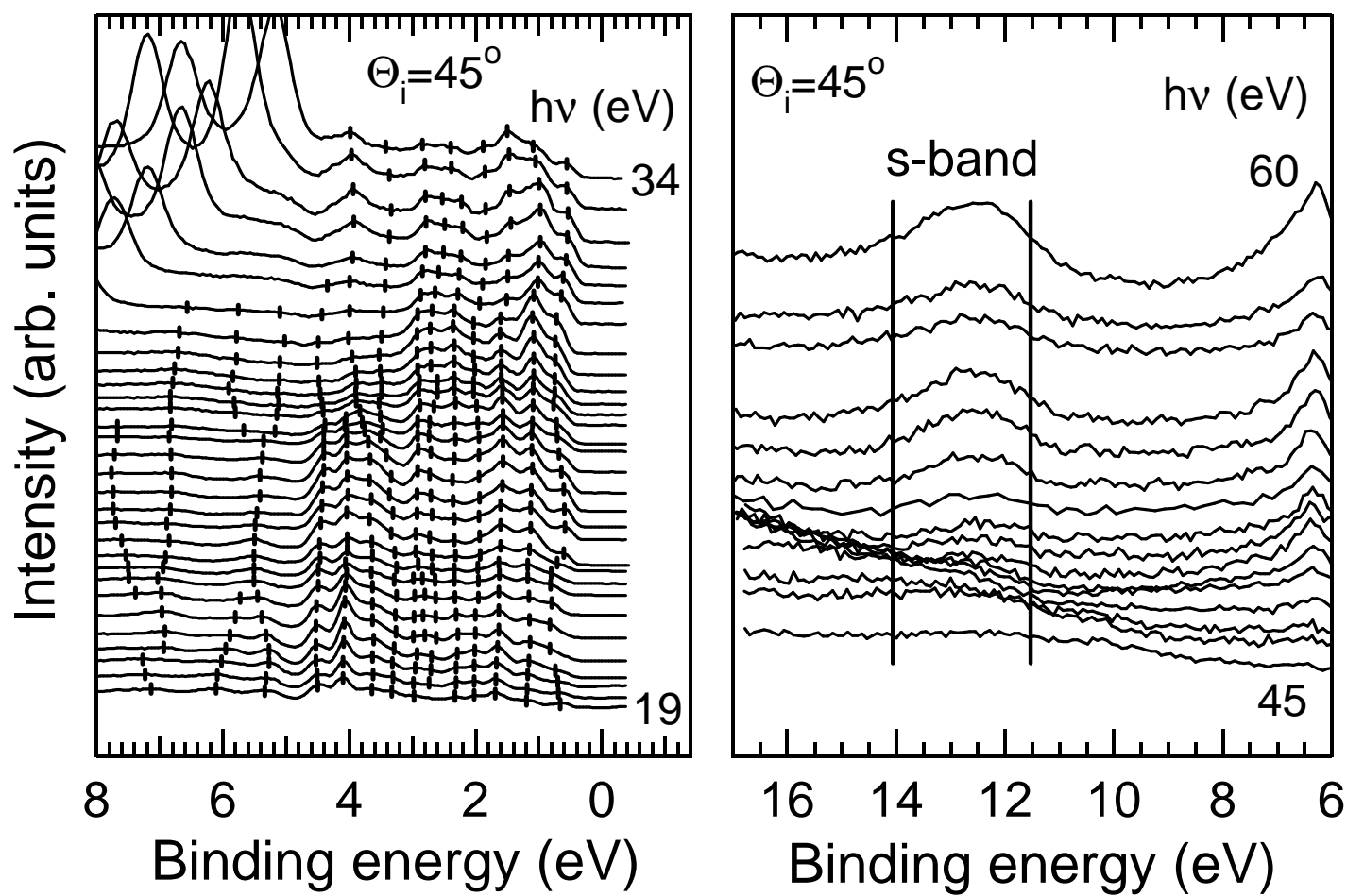
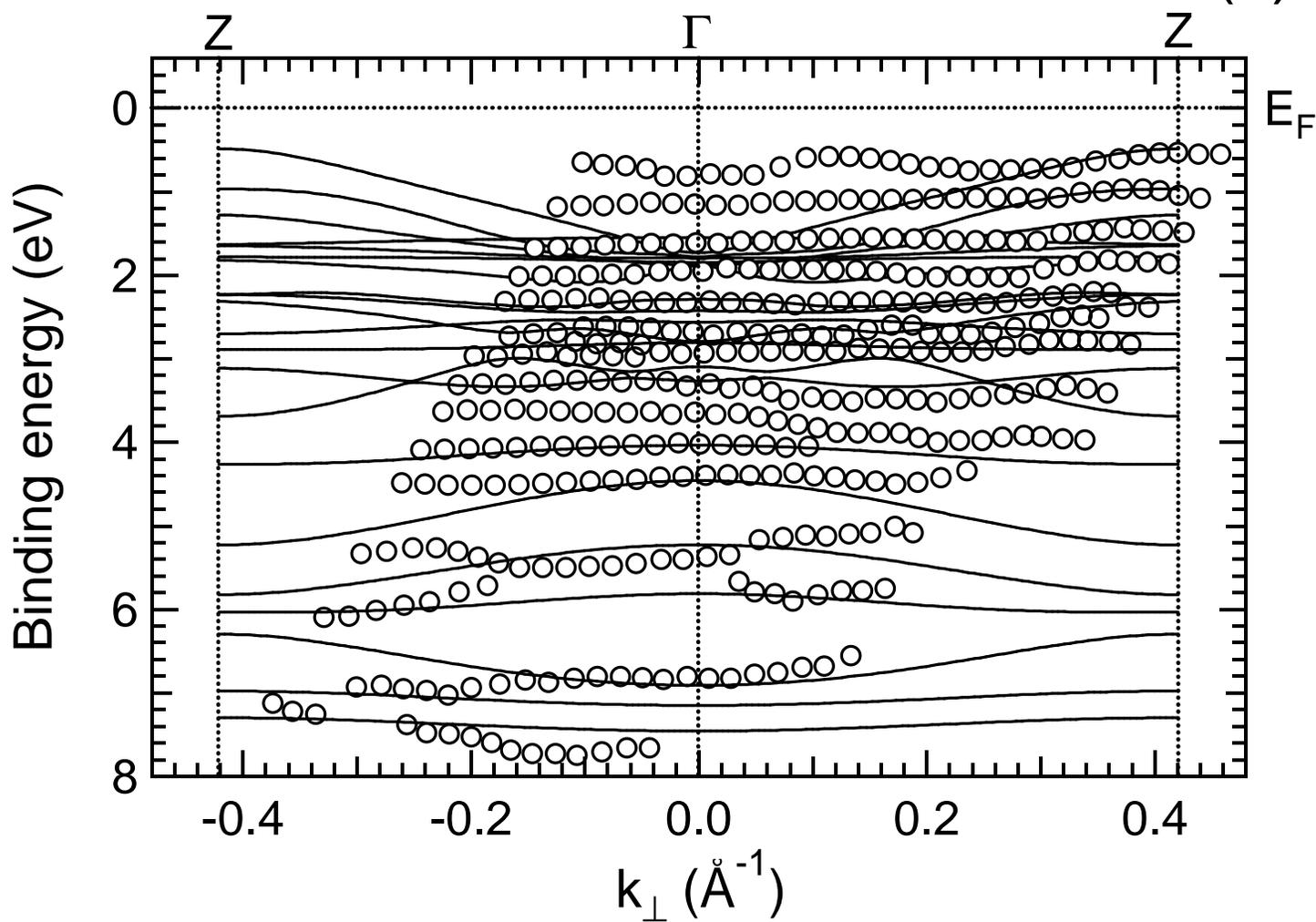
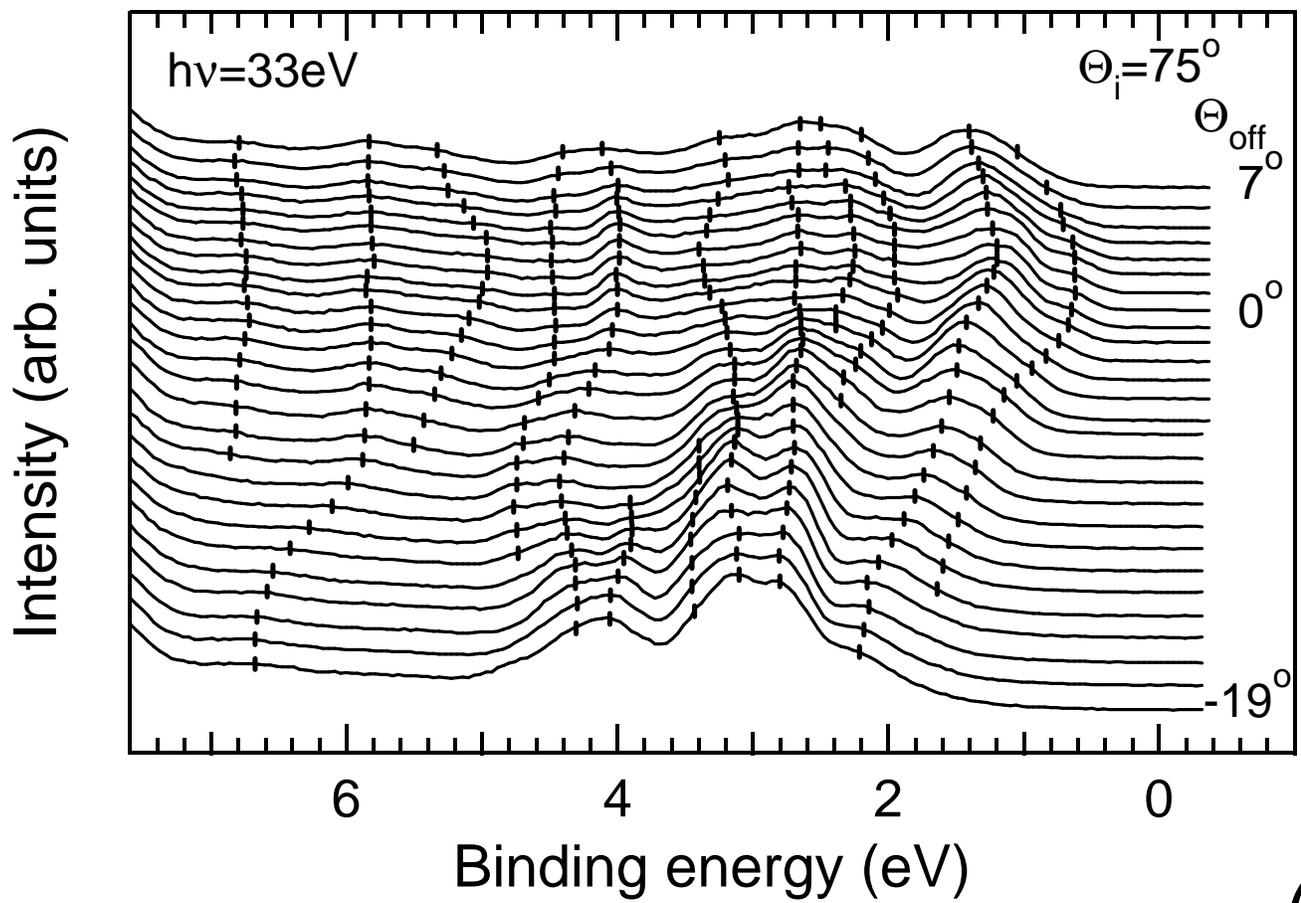
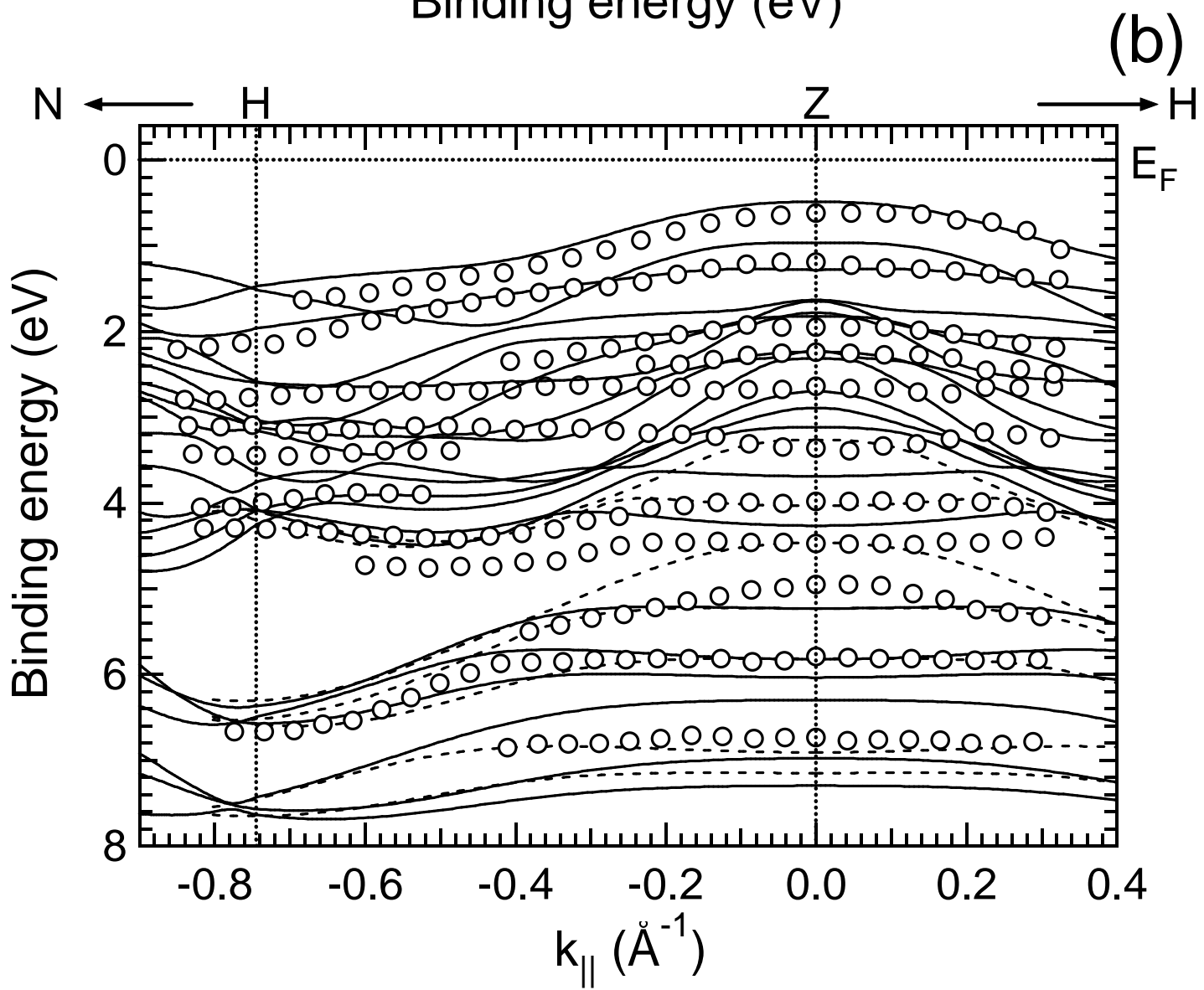
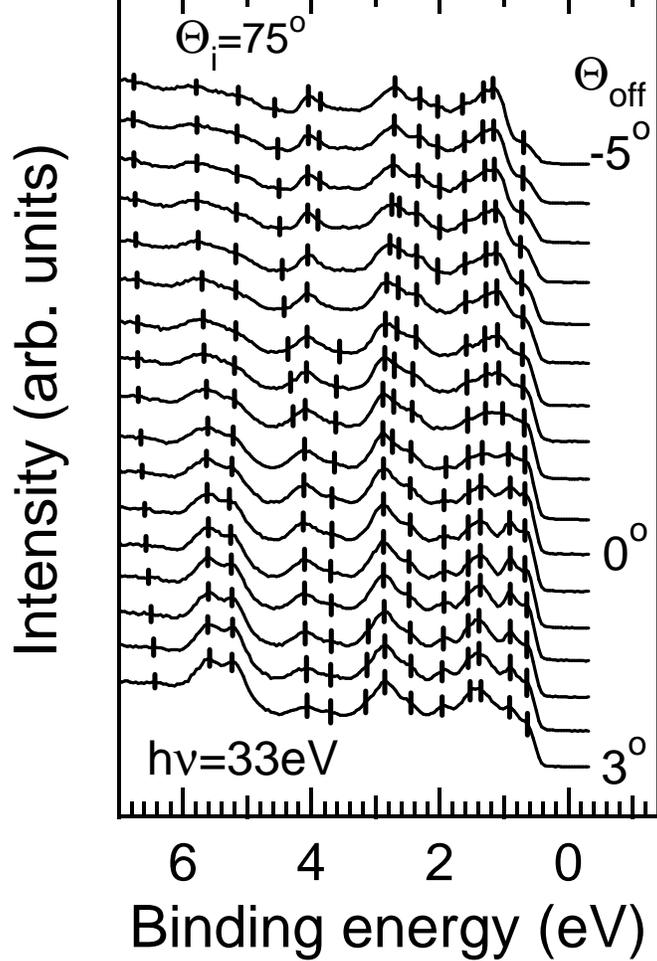
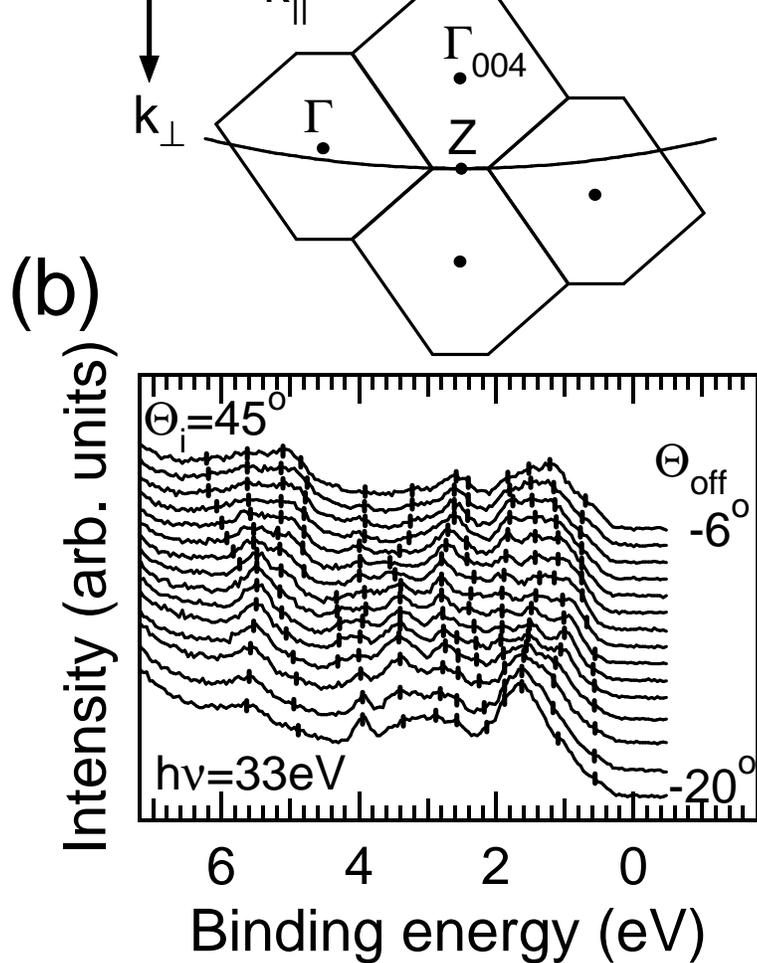
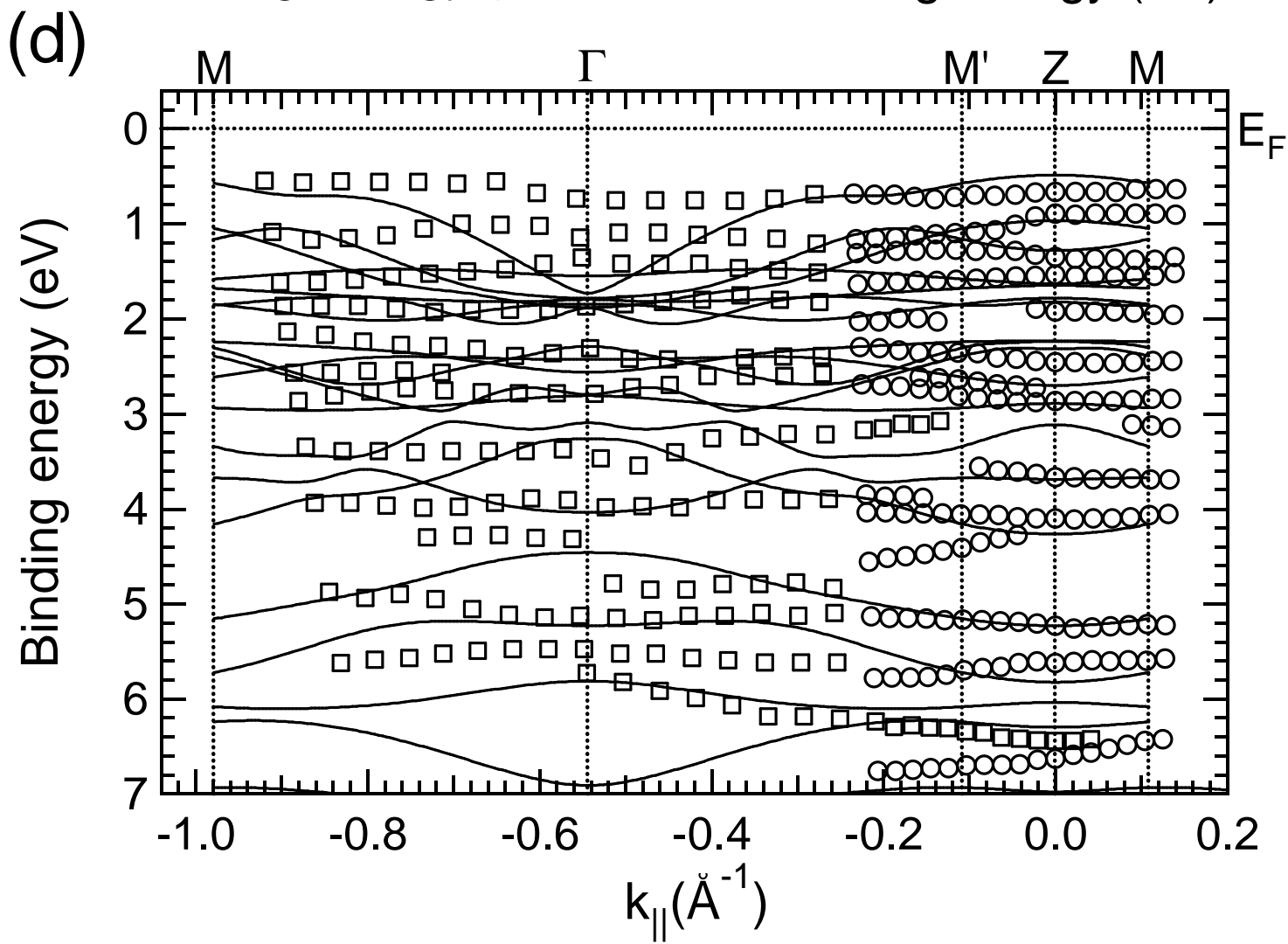
\end{document}